\documentstyle[12pt]{article}
\begin{document}
\def\t{\times}
\def\p{\phi}
\def\P{\Phi}
\def\a{\alpha}
\def\e{\varepsilon}
\def\be{\begin{equation}}
\def\ee{\end{equation}}
\def\l{\label}
\def\0{\setcounter{equation}{0}}
\def\b{\beta}
\def\S{\Sigma}
\def\C{\cite}
\def\r{\ref}
\def\ba{\begin{eqnarray}}
\def\ea{\end{eqnarray}}
\def\n{\nonumber}
\def\R{\rho}
\def\X{\Xi}
\def\x{\xi}
\def\La{\Lambda}
\def\la{\lambda}
\def\d{\delta}
\def\s{\sigma}
\def\f{\frac}
\def\D{\Delta}
\def\pa{\partial}
\def\Th{\Theta}
\def\o{\omega}
\def\O{\Omega}
\def\th{\theta}
\def\ga{\gamma}
\def\Ga{\Gamma}
\def\h{\hat}
\def\rar{\rightarrow}
\def\vp{\varphi}
\def\inf{\infty}
\def\le{\left}
\def\ri{\right}
\def\foot{\footnote}
\def\u{\underline}

\begin{center}
{\Large\bf Fields topology and perturbation theory}
\vskip 0.2cm

{\large\it J.Manjavidze\\{JINR, Dubna, Russia.\\
Inst. of Physics, Tbilisi, Georgia}}
\end{center}
%\huge

\begin{abstract}
{\footnotesize The fields nonlinear modes quantization scheme is
discussed.  New form of the perturbation theory achieved by unitary
mapping the quantum dynamics in the space $W_G$ of ({\it action,
angle})-type collective variables.  It is shown why the transformed
perturbation theory contributions may accumulated exactly on the
boundary $\pa W_G$.  Abilities of the developed formalism are
illustrated by examples from quantum mechanics and field theory.}
\end{abstract}

$\bullet$ {\it Introduction}

One may use for my talk another titles. For instance:

-- {\bf Quantization of nonlinear modes (waves)}

-- {\bf Strong-coupling perturbation theory}

-- {\bf Examples of unitary transformation of path-integral variables}

-- {\bf Quantum theory and symplectic geometry}

-- {\bf Path-Integral solution of H-atom Problem}

-- {\bf Particles Creation in the Integrable Systems}

-- {\bf Symmetry breaking in the $O(4,2)$-invariant field theories}

-- {\bf Ghost-free quantization of Yang-Mills fields}
\\
But the role of topologies in the perturbation theory structure
appears so new for me that I offer concentrate the attention just
on this question. (Details of the formalism one can find in
hep-th/9811160.)

We will consider the expansion of integral:
$$
A(F)=\int Du e^{iS(u)}F(u)
$$
in vicinity of $real-time$ path
$$
u_c~:~\f{\d S(u_c)}{\d u_c}\equiv 0.
$$
This demands knowledge of Green function $G$. But the equation:
$$
(\pa^2_\mu+v''(u_c))_xG(x,x';u_c)=\d(x-x').
$$
is translationally noninvariant if $u_c=u_c(x)$ and therefore
has not explicit solution. By this reason this ordinary (WKB) method
has not a future.

I would like to demonstrate the strict {\it strong-coup\-ling}
perturbation theory, $u_c=O(1/g)$.  Actually I will follow the idea
that {\it the substitution may considerably simplify calculations},
helping avoid above problem.

$\bullet$ {\it Unitarity condition}

Let us calculate probability to find somewhere a particle with energy
$E$
$$
\R_1(E)=\int dx_1 dx_2 A_1(x_1 ,x_2 ;E)A^*_1 (x_1 ,x_2 ;E).
$$
using the spectral representation of one-particle amplitude:
$$
A_1(x_1 ,x_2 ;E)=\sum_{n}\frac{\Psi^*_{n} (x_2)\Psi_n (x_1)}{E-E_n
+i\e}, ~~~\e \rar +0,
$$
All unnecessary contributions with $E\neq E_n$ were canceled in
$\R_1(E)$:
\be
\R_1(E)=\sum_{n}|\frac{1}{E-E_n+i\e}|^2 =\frac{\pi}{\e}\sum_{n}\d
(E-E_{n}),
\l{1}\ee
together with real part of propagator $1/E-E_n+i\e$.

{\bf I would like to exclude such unnecessary contributions} (like
$E\neq E_n$). This means inclusion of last equality in (\r{1}) into
formalism, i.e. transition from $A_1A_1^*$ to absorption part $Im
A_1$.

Then the unitarity (optical theorem: $2i\e|A_1|^2=A_1-A^*_1$) becomes
sufficient and necessary condition: I will show, it defines the
$complete$ set of contributions in the physically acceptable domains.

The formalism based on the statements:\\
{\bf (A)} If {$A\sim e^{iS(u)}$} and $S(u)$ is the action, if
$\R\sim|A|^2$ is measurable, if the equality $2i\e|A|^2=A-A^*$ is
taken into account, then:  $$ \R(F)= e^{-i\h{K}(ej)}\int
DM(u,j) e^{-iU(u,e)}\tilde{F}(u,e) \equiv \h{\cal
O}(u)\tilde{F}(u) $$ where expansion over differential operator $$
\h{K}(ej)=\f{1}{2}\int dx\f{\d}{\d
j(x)}\f{\d}{\d e(x)}\equiv\f{1}{2}\int dx \h{j}(x)\h{e}(x)
$$
generates the perturbation series, functional
$U(u,e)=O(e^3)$ describes interactions, the measure $DM$ is Diracian
($\d$-like):
$$
DM=\prod_x du(x)\d\le(\f{\d S(u)}{\d u(x)}+j(x)\ri).
$$
At the very end one should take $j=e=0$.

$Note$: the variational principle is $derivable$ from above first
quantum principles.

{\bf (B)} If coordinate $u_c(\x,\eta)$ and corresponding
momentum $p_c(\x,\eta)$ obey the equations
\be
\{u_c,h_j\}=\f{\pa H_j}{\pa
p_c},~\{p_c,h_j\}=-\f{\pa H_j}{\pa u_c},~{\rm at~arbitrary}~j,
\l{2}\ee
where $H_j(u,p)=\f{1}{2}p^2+v(u)-ju$ and $\{,\}$ is the Poisson
bracket, if
\be
h_j(\x,\eta)=H_j(u_c,p_c),~h(\eta)\equiv h_0(\x,\eta),
\l{3}\ee
then:

(a) the transformed measure has the form:
$$
DM(\x,\eta)=\prod_t d\x(t)\eta(t)\d(\dot{\x}-\f{\pa
h_j}{\pa\eta}) \d(\dot{\eta}+\f{\pa h_j}{\pa\x}),
$$
(it is $T\rar W_G$ mapping) since, as follows from (\r{2}, \r{3})

(a') $(u_c,p_c)$ are the solutions of incident (classical)
Hamiltonian equations:
$$
\{u_c^i,u_c^k\}=\{p_c^i,p_c^k\}=0,~~\{u_c^i,p_c^k\}=\d^{ik}
$$

(b) ${\rm dim}W_G\leq{\rm dim}T$, where $T$ is the incident phase
space (reduction of $T$).

(c) ${\rm dim}W_G$ may be even or odd (splitting: $W_G=T^*G\t R$).

{\bf (C)} If the Green function $g(t-t')$ of equations
$$
\dot{\x}=\f{\pa h_j}{\pa\eta},~\dot{\eta}=-\f{\pa h_j}{\pa\x},
$$
$h_j(\x,\eta)=H_j(u_c,p_c)$, have the form:
$$
g(t-t')=\th(t-t'),~g(0)=1,
$$
then the quantum corrections to semiclassical approximation are
accumulated on the boundaries of $T^*G$:
$$
\R =\R^{sc}+\int_{\pa T^*G}d\R^{q},
$$
where $\R^{sc}$ is the semiclassical contribution. The explicit form of
quantum corrections term $d\R^{q}$ will be given.

$\bullet$ The generalization of formalism on the field theory, where
$u_c=u_c(\vec{x};\x,\eta)$, becomes evident noting (b) and
considering space coordinates as the indexes of special cell.

By the same reason ((b) and (c)) the formalism allows to consider also
the situation where $(\x,\eta)=(\x,\eta)(x,t)$. Last one incorporates
the gauge freedom.

{\bf (D)} The measure
$$
DM(\x,\eta)=\prod_t d\x(t)d\eta(t)\d(\dot{\x}-\o(\eta)-j\f{\pa
u_c}{\pa\eta}) \d(\dot{\eta}+j\f{\pa u_c}{\pa\x}),
$$ admits the cotangent foliation of quantum force $j$:
$$
DM(\x,\eta)=\prod_t d\x(t)\eta(t)\d(\dot{\x}-\o(\eta)-j_\x)
\d(\dot{\eta}-j_\eta)
$$
$$
\h{K}(je)=\f{1}{2}\int dt\{\h{j}_\x\h{e}_\x+\h{j}_\eta\h{e}_\eta\}
$$
and
$$
e\rar e_c=e_\x\f{\pa u_c}{\pa\eta}-e_\eta\f{\pa u_c}{\pa\x}
$$

-- Cotangent foliation of quantum force $j$ gives the completely
Hamiltonian description

-- The foliation allows quantize all classical degrees of freedom
independently

-- The auxiliary variable $e_c$ is invariant of canonical
transformations

-- The perturbation theory describes fluctuations of classical flow
through elementary cell $\d u\wedge\d p$ in the $T^*G$ cotangent
(sub)space

-- The foliation solves the technical problem of functional
determinants calculation.

\begin{center}
{\rm\bf Content}
\end{center}

$\bullet$ \underline{\it Introduction into formalism}

-- The representation $\R(F)= \h{\cal O}(u)\tilde{F}(u)$ is derived.

-- The mechanism of canonical and coordinate transformations is
shown.

$\bullet$ \underline{\it Description of the perturbation theory}

-- The main properties of new perturbation theory are shown using
simplest quantum-mechanical example.

$\bullet$ \underline{\it Theory of transformation}

-- The Coulomb problem is solved.

-- The $\R$ is calculated for sin-Gordon model in the space of
solitons parameters. The consequences are discussed.

-- The quantum measure of the scalar $O(4,2)$-invariant field theory
in the $W_O=O(4,2)/O(4)\t O(2)$ factor space is derived. It is shown
that the scale invariance is broken.

-- The measure of Yang-Mills theory in the $W_O\t G$ space, where $G$
is the gauge group, is derived.

\begin{center}
{\rm\bf Definitions}
\end{center}

$\bullet$ {\it S-matrix theory}

The $n$-into-$m$ particles transition amplitudes is
$$
a_{nm} (q';q)
=\prod^{m}_{k=1}\h{\p}(q'_k)\prod^{n}_{k=1}\h{\p}^* (q_k)Z(\p),
$$
where $q'_k$ ($q_k$) are the in- (out)-going particles momenta. The
`hat' symbol means:
$$
\h{\p}(q) \equiv \int dx e^{-iqx} \f{\d}{\d
\p (x)} \equiv \int dx e^{-iqx} \h{\p}(x).
$$

The vacuum-in-vacuum transition amplitude in the auxiliary background
field $\p$ is
\be
Z(\p)=\int Du e^{iS_0 (u) -iV(u+\p)},
\l{4}\ee
where $S_0$ is the free part of action:
$$
S_0 (u)=\f{1}{2}\int_{C_+} dx ((\pa_{\mu}u)^2 - m^2 u^2)
$$
and $V$ describes the interactions:
$$
V(u)=\int_{C_+} dx v(u).
$$
The time integrals in (\r{4}) are defined on Mills time contour:
$$
C_+ :t \rar t+i\e,~~~\e \rar +0,~~~-\infty \leq t \leq +\infty.
$$

$\bullet$ {\it S-matrix interpretation of statistics}

Let us calculate now the probability
$$
\tilde{\R}_{nm}(P)=\f {1}{n!m!}\int d\o_m (q') d\o_n (q)
\d (P-\sum^{m}_{k=1}q'_k)\d (P-\sum^{n}_{k=1}q_k)|a_{nm}|^2,
$$

Summation over all $n,m$ gives the generating functional:
$$
\R(\a,z)=e^{-N_+(\a_+,z_+;\h\p)-
N_-(\a_-,z_-;\h\p)}\R_0(\p)
\equiv e^{- N(\a,z;\h\p)}\R_0(\p).
$$
in the Fourier-Mellin representation.

The external particles number operator
$$
N_{\pm}(\a,z;\h\p)\equiv \int d\o_1 (q)e^{-iq\a_\pm}z_\pm(q)
\h \p_{\pm}(q)\h \p^*_{\mp}(q).
$$
and
$$
\R_0 (\phi)=Z(\phi_+)Z^*(-\phi_-).
$$
describes the vacuum-into-vacuum transition.

\begin{center}
{\it Comments}
\end{center}

$\bullet$
$\R(\a,z)$ may be used for generation of cross sections helping
Fourier-Mellin transformation.

$\bullet$
If (i) $\a_\pm=(i\b_\pm,\vec{0})$, where
$$
\bar\b_\pm: E_\pm\equiv-\f{\pa}{\pa\bar\b_\pm}\ln\R(\bar\b_\pm),
$$
and if (ii) the fluctuations near $\bar\b_\pm$ are Gaussian, then
$\R(\bar\b,z)$ is the grand partition function. Then $\bar\b_\pm$ is
the inverse temperature in the initial (final) state and $z(q)$ is
the activity.

$\bullet$
Including the black-body environment $\R(\bar\b,z)$ is the
generating functional of the real-time finite temperature field
theory of Schwinger-Keldysh type. If there is not correlations on
the time infinities, then $\R(\bar\b,z)$ may be continued on the
Matsubara imaginary-time contour.

$\bullet$
$S$-matrix interpretation allows to extend the formalism on the
nonequilibrium media considering $\b_\pm =\b_\pm(Y)$, where $Y$ is the
4-coordinate of {\it measurement}.

\begin{center}
{\bf Unitary definition of measure}
\end{center}

$\bullet$ {\it Factorization property}

In the expression
$$
\R(\a,z)=e^{- N(\a,z;\h\p)}\R_0(\p).
$$
the `external' properties, fixed by $\a,~z$, and `internal' ones,
described by $\R_0(\p)$, are {\it factorized}: the operator $e^{-
N(\a,z;\h\p)}$, where $N=N_++N_-$ and
$$
N_{\pm}(\a,z;\h\p)\equiv \int d\o_1 (q)e^{-iq\a_\pm}z_\pm(q)
\h \p_{\pm}(q)\h \p^*_{\mp}(q),
$$
maps interacting fields system,
$$
\R_0(\phi)=\int Du_+ Du_- e^{iS_0 (u_+) -iU(u_+ +\phi_+)}
e^{-iS_0 (u_-) +iV(u_- -\phi_-)},
$$
on the observable states.

This property reflects adiabaticity of quantum perturbations.

$\bullet$ {\it Dirac measure}

We will use the substitution :
$$
u(x)_{\pm}=u(x)\pm\vp(x),~
$$
with boundary conditions:
$$
\int_{\s_{\infty}} dx_{\mu}\vp(x)\pa^{\mu}u(x)=0~:~\vp(x)|_{(x)\in
\s_{\infty}}=0
$$
to establish the equality:
$
2iAA^*=A-A^*.
$
Expanding over $\vp$:
$$
\R_0(\phi)=e^{-i\h{K}(j\vp)}\int DM(u,j)
e^{-iU(u,\vp)}
e^{i2\Re\int_{C_+}dx \p(x)(j(x)-v'(u))}
$$
where
$$
DM(u)=\prod_{x} du(x) \d (\pa_{\mu}^2 u +
m^2 u +v'(u)  -j),
$$
$$
\h{K}(j,\vp)=\f{1}{2}\Re\int_{C_+} dx \hat{j}(x) \hat{\vp}(x),
$$
$$
U(u,\vp)=V(u+\vp) - V(u- \vp)-
2\Re\int_{C_+} dx \vp (x)v'(u)=O(\vp^3).
$$
$Note$: last exponent is linear over $u(x)$ because of $\d$-likeness
of measure $DM$.

$\bullet$ {\it Generating functional}

Substitution of $\R_0$ gives {\bf (A)}:
$$
\R(\b,z)=e^{-i\h{K}}\int DM(u)e^{+iS_0(u)-iU(u;\vp)}e^{N(\b,z;u)},
$$
where
$$
N(\b,z;u)=n(\b_+,z_+;u)+n^*(\b_-,z_-;u)
$$
and
$$
n(\b,z;u)=\int d\o (q) e^{-\b (\e(q) +\mu (q))}
|\Ga(q,u)|^2,
$$
where $\mu (q)=\ln z(q)/\b$ is the chemical potential and
$$
\Ga (q,u)=\int_{C_+}dx e^{iqx} (\pa^2 +m^2)u(x)
$$$$
DM(u)=\prod_{x} du(x) \d (\pa_{\mu}^2 u +
m^2 u +v'(u)  -j),
$$
$$
\h{K}(j,\vp)=\f{1}{2}\Re\int_{C_+} dx \hat{j}(x) \hat{\vp}(x),
$$
$$
U(u,\vp)=V(u+\vp) - V(u- \vp)-
2\Re\int_{C_+} dx \vp (x)v'(u)=O(\vp^3).
$$
$Note$: if $u_c(x)$ ia a `good' function, then $\Ga (q,u)=0$ since
$q^2=m^2$ by definition.

\begin{center}
{\it Comments}
\end{center}
{\bf a.}The functional $\delta$-function is defined as follows:
$$
\prod_x
\delta (f_x(u))=\int \prod_{x}\frac{de(x)}{\pi}
e^{-2i\Re\int_{C_+}dx e(x)f_x(u)}
$$
So, considering the double integral $AA^*$ we may introduce
integration over two independent fields $u$ and $e$. Then, (i)
integral over $e$ gives the $\d$-function and (ii) last one defines
integral over $u$. One can say: the real-time theories
are `simple'.

It should be underlined that the measure $DM$ was derived for
$real-time$ processes only.

{\bf b.} Only $strict$ solutions of equation
\be
\pa_\mu^2 u+v'(u)=0,
\l{5}\ee
should be taken into account.

{\bf c.} $\R$ is described by the $sum$ of all solutions of
eq.(\r{5}), independently from theirs `nearness' in the functional
space.

{\bf d.} $\R$ did not contain the interference terms from various
topologically nonequivalent contributions. This displays the
orthogonality of corresponding Hilbert spaces.

{\bf e.} The measure $DM$ and $\h{K}$ includes $j(x)$ as the external
source.  Its fluctuation disturb the solutions of eq.(\r{5}).

{\bf f.} If $j$ is switched on adiabatically then the field
disturbed by $j(x)$ belongs to the same manifold (topology class) as
the classical field $u_c$.

{\bf g.} ({\it Selection rule}) If $V_{u_c}$ is the  zero-modes
volume occupied by given $u_c$, then taking into account {\bf b, c}
and {\bf f} one should leave the contribution with largest $V_{u_c}$:
if ${\rm dim} V_{u_c}> {\rm dim}V_{u'_c}$, then $u'_c$ contributions
may be neglected with $O(V_{u'_c}/V_{u_c})$ accuracy.

$Note$: the imaginary-time (kink-like) contributions may be neglected
iff theirs contribution are realized on zero ($V_{u'_c}/V_{u_c}=0$)
measure.

{\bf h.} Our definition of $\R$ restores the stationary phase methods
perturbation theory in the vicinity of trivial extremum $u_c=0$. The
comparison of our and WKB perturbation theories is impossible for
the case $u_c\neq0$ since last one is unknown for this case.

{\bf g.} The $i\e$-prescription should be used to avoid the
singularities and for right definition of time analytical
continuation to connect (if this is not in contradiction with
topological principles) the real- and imaginary-time trajectories.

\begin{center}
{\bf Canonical transformation}
\end{center}

$\bullet$ {\it Introduction into the transformation theory}

Let's start consideration from (1+0) field theory
$$
A_1(x_1 ,x_2;E)=i\int^{\infty}_{0}dTe^{iET}
\int\prod dx(t)dp(t)e^{iS_{C_+}(x,p)},
$$
assuming that $x(0)=x_1$, $x(T)=x_2$. Then
$$
DM(x,p)=\delta (E-H_T)
\prod_{t}dx dp
\d(\dot{x}-\f{\pa H_j}{\pa p})\d(\dot{p}+\f{\pa H_j}{\pa x}),
$$
where
$$
H_{j}=\frac{1}{2}p^2 +v(x)-jx,~H_T=H_{j=0}|_{t=T}.
$$
Inserting
$$
1=\int D\th Dh\prod_{t}\d (h-\frac{1}{2}p^2 -v(x))
\n\\\t
\d (\th-\int^{x}dx(2(h-v(x)))^{-1/2}).
$$
If: $h_j(\th,h)=H_j(x_c,p_c)=h-jx_c(\th,h)$, then
$$
DM(\th ,h)=\d (E-h(T))\prod_{t}d\th dh\d (\dot{\th}-\frac{\pa h_j}
{\pa h}) \d(\dot{h}+\f{\pa h_j}{\pa\th}),
$$

$\bullet$ $j\pa x_c(\th,h)/\pa h$ and $j\pa x_c(\th,h)/\pa\th$ in
eqs.:
$$
\dot{\th}=\f{\pa h_j}{\pa h}=1-j\frac{\pa x_c(\th,h)}{\pa h},~
\dot{h}=-\f{\pa h_j}{\pa\th}=j\f{\pa x_c(\th,h)}{\pa\th}.
$$
are the projections of $j$ on the axis of $W=(\th,h)$ space. Using
identity (at $j_b=e_b=0$):
$$
\prod\d(a\pm bj)=e^{\a^{-1}\int \h{j}_b\h{e}_b}e^{\mp\a\int jbe_b}
\prod\d(a-j_b)
$$
($\a$ is arbitrary) one can complete the mapping:
$$
DM=\d(E-h_T)\prod_t d\th dh\d(\dot{\th}-1-j_\th) \d(\dot{h}-j_h),
$$$$
\h{K}(j\cdot e)=\f{1}{2}\Re\int_{C_+}dt
\{\h{j}_\th\h{e}_\th+\h{j}_h\h{e}_h\},
$$$$
e\rar e_c=e_\th\frac{\pa x_c(\th,h)}{\pa h}-e_h\f{\pa
x_c(\th,h)}{\pa\th}.
$$
Action of $\exp\{-i\h{K}(j\cdot e)\}$ gives:
$$
\R(E)=\int^\infty_0 dT \int DM :e^{-U(\h{e}_c,x_c)}e^{iS_0(x_c)}:,
$$$$\h{e}_c=\{\h{j}\wedge\h{W}\}x_c,~j=(j_\th,j_h),~W=(\th,h).
$$
This completes the (a) part of {\bf (B)}.

$\bullet$ {\it Zero modes}

Noting that
$$\int \prod_t dX(t)\d(\dot X)=\int dX(0)=V_X
$$
the measure
$$DM=\d(E-h_T)\prod_t d\th dh\d(\dot{\th}-1-j_\th) \d(\dot{h}-j_h)
\sim d\th(0)
$$This is the translational zero modes measure.

\begin{center}
{\it Comments}
\end{center}

$\bullet$

-- The zero modes differential measure was defined without
Faddeev-Popov ansatz.

-- The Faddeev-Popov ghosts would not arise.

$\bullet$

-- The ghost-free quantization scheme may be shown for Yang-Mills
field theories.

-- This removes (?) the problem of Gribov's ambiguities.

$\bullet$  {\it  Perturbation theory structure}
$$\R(E)=\int^\infty_0 dT \int DM :e^{-U(\h{e}_c,x_c)}e^{iS_0(x_c)}:,
$$$$
DM=\d(E-h_T)\prod_t d\th dh\d(\dot{\th}-1-j_\th) \d(\dot{h}-j_h),
$$
Let $g(t-t')$ be the Green function. The $i\e$-prescription
gives:
$$
g(t-t')=\th(t-t'),~g(0)=1,~g(t-t')g(t'-t)=0,~
$$$$
g^2(t'-t)=g(t'-t),~g(t'-t)+g(t-t')=1.
$$
$$
e^{-iU (x_c,\h{e}_c )}=\prod^{\infty}_{n=1}\prod^{2n+1}_{k=0}
e^{-iV_{k,n}(\h{j},x_c)},
$$
where
$$
V_{k,n}(\h{j},x_c)=\int^{T}_{0}dt (\h{j}_{\p}(t))^{2n-k+1}
(\h{j}_{I}(t))^{k}b_{k,n}(x_c).
$$
The explicit form of $b_{k,n}(x_c)$ is not important.
$$
\h j_X(t)=\int dt' g(t-t')\h X(t'),~X=(\x,\eta)
$$
$$ \h{j}_X(t_1)b_{k,n}(x_c(t_2))=\Th (t_1 -t_2)
\pa b_{k,n}(x_c)/\pa X_0
$$
since $x_c =x_c(X(t)+X_0)$, or
$$
\h{j}_{X,1}b_2=\Th_{12}\pa_{X_0}b_2
$$
since indices $(k,n)$ are not important.

$k=0,m=1$
$$
\h{j}_1b_1 = \Th_{11}\pa_0 b_1 =\pa_0 b_1 \neq 0.
$$
$k=0,m=2$
$$
\h{j}_1 \h{j}_2 b_1 b_2
=\Th_{21} b^2_1 b_2 + b^1_1 b^1_2 +\Th_{12}b_1 b^2_2,
$$
($b^n_i \equiv \pa^n b_i$). Inserting $1=\Th_{12}+\Th_{21}$:
$$
\h{j}_1 \h{j}_2 b_1 b_2 =\Th_{21}( b^2_1 b_2 + b^1_1 b^1_2)+
\Th_{12}(b_1 b^2_2+b^1_1 b^1_2) \n \\ =
$$$$
=\pa_0 (\Th_{21} b^1_1 b_2+\Th_{12}b_1 b^1_2)
$$
This important property of the perturbation theory is conserved in
arbitrary order over $m$ and $k$.

This ends the statement {\bf (C)}:
$$
\R=\R^{sc}+\int_{\pa W_G}d\R^{q}.
$$

$\bullet$ {\it General theory of transformation}

If $J_i=J_i(x,p)$, $i=1,2,...,N$, are the first integrals in
involution then the equations
$$
\dot{J}=-\f{\pa{\cal H}}{\pa Q},~\dot{Q}=\f{\pa{\cal H}}{\pa J},
$$
\be
\eta=J(x,p),~\x=Q(x,p)
\l{6}\ee
solves mechanical problem (Liouville-Arnold).

Corresponding mapping:
$$
J:T \rar W_G,
$$introduces integral $manifold$ $J_{\o}=J^{-1}(\o)$ to which the
$classical$ phase flaw belongs \underline{$completely$}.

{\bf Suggestion A}: {\it If we know the classical phase flow
$(x,p)_c$, then
(i) one can restore $W_G$ without pointing out the canonical
mapping (\r{6});
(ii) quantum dynamics representations in $(\x,\eta)\in T^*G$ and
$(x,p)\in T$ are isomorphic.}

-- This assumes following substitution:
$$
1=\f{1}{\D}\int \prod_t d\x d\eta\d(u(t)_x-u_c(\x,\eta)_x)
\d(p(t)_x-p_c(\x,\eta)_x),
$$
where $(u,p)_c$ obey the {\bf (B)} conditions.

\begin{center}
{\it Comments}
\end{center}

$\bullet$ It was noted: ${\rm dim}W_G\leq{\rm dim}T$; ${\rm dim}W_G$
may be even or odd.

I wish demonstrate the reduction:
$$
T^4\rar W_C^3=T^*G^2\t R^1
$$
considering the {\bf Coulomb problem}:
$$r_c=\f{\eta_1^2(\eta_1^2+\eta_2^2)^{1/2}}
{(\eta_1^2+\eta_2^2)^{1/2}+\eta_2\cos \x_1},~\vp_c=\x_1,
p_c=\f{\eta_2\sin \x_1}{\eta_1(\eta_1^2+\eta_2^2)^{1/2}},~l_c=\eta_1.
$$

$$
U_T(r,e)=-S_0(r)+
+\int^T_0 dt[\f{1}{((r+e_r)^2+r^2e_{\vp}^2)^{1/2}}
-\f{1}{((r-e_r)^2+r^2e_{\vp}^2)^{1/2}}+2\f{e_r}{r}]
$$
$$
e_r=e_{\eta_1}\f{\pa r_c}{\pa \x_1}-e_{\x_1}\f{\pa r_c}{\pa \eta_1}.
$$

$$
DM(\x, \eta)=\d(E-h(T))\prod_td^2\x d^2\eta
\d(\dot{\x}_1-\o_1-j_{\x_1})
$$$$\t
\d(\dot{\x}_2-\o_2-j_{\x_2})
\d(\dot{\eta}_1-j_{\eta_1})
\d(\dot{\eta}_2),~\o_i=\pa h/\pa\eta_i.
$$

We have put $j_{\x_2}=e_{\x_2}=0$ since $r_c= r_c (\x_1, \eta_1,
\eta_2)$:
$$
\h{K}(j,e)=\int^T_0 dt (\h{j}_{\x_1}\h{e}_{\x_1}+
\h{j}_{\eta_1}\h{e}_{\eta_1}),
$$

Using last $\d$-functions:
$$
\R(E)=\int^\infty_0 dT e^{-i\h{K}(j,e)}\int dM
e^{-iU_T(r_c,e)}e^{iS_0},
$$
where
$$
dM=\f{d\x_1
d\eta_1}{\o_2(E)}
$$$$
r_c(t)=r_c(\eta_1 +\eta(t),
\bar{\eta}_2(E,T), \x_1+\o_1(t) +\x(t)).
$$

The integration range over $\x_1$ and $\eta_1$ is as follows:
$$\pa W_C: 0\leq \x_1 \leq 2\pi,~~-\infty \leq \eta_1 \leq +\infty.
$$
Then,
$$
\R(E)=\int^\infty_0 dT \int dM \{e^{iS_0 (r_c)}+ \f{\pa}{\pa
\x_1}b_{\x_1}+\f{\pa}{\pa\eta_1}b_{\eta_1}\}.
$$
and the mean value of quantum corrections in the $\x_1$
direction are equal to zero:
$$
\int^{2\pi}_0 d\x_1 \f{\pa}{\pa \x_1}b_{\x_1}(\x_1,
\eta_1) =0
$$
since $r_c$ is the closed trajectory independently from initial
conditions.

In the $\eta_1$ direction the motion is classical:
$$
\int^{+\infty}_{-\infty} d\eta_1 \f{\pa}{\pa \eta_1}
b_{\eta_1}(\x_1, \eta_1)=0
$$
since (i) $b_{\eta_1}$ is the series over $1/r_c^2$ and (ii) $r_c
\rar \infty$ when $|\eta_1| \rar \infty$.

This is the desired result:
$$
\R(E)=\int^\infty_0 dT \int dM e^{iS_0 (r_c)}.
$$
Noting that
$$
S_0 (r_c)= kS_1 (E),~~k=\pm 1, \pm 2,...,
$$
where $S_1 (E)$ is the action over one classical period $T_1$:
$$
\frac{\partial S_1 (E)}{\partial E}=T_1 (E),
$$
and using the identity:
$$
\sum^{+\infty}_{-\infty} e^{inS_1 (E)} =
2\pi \sum^{+\infty}_{-\infty}\d (S_1 (E) - 2\pi n),
$$
we find normalizing on zero-modes volume, that
$$
\R(E)=\pi \sum_{n} \d (E + 1/2n^2).
$$

{\bf Suggestion B}: {\it
If $v(u)$ interaction potential, if $v_X(u_c)$ is the derivative of
$v(u_c)$ in the $X=(\x,\eta)$ direction, if $\{v_X(u_c)\}$ is
corresponding manifold, then the theory is exactly semiclassical iff
$$
\{v_X(u_c)\}\bigcap\pa_X T^*G=\emptyset,~X(\x,\eta)\in T^*G,
$$
and $W_G=T^*G\t R$, $T^*G\neq\emptyset$.}

$\bullet$ The mapping
$$
T(\infty)\rar W_{sG}=T^*G(2N),~N=1,2,...
$$
may be investigated considering (1,1)-dimensional {\bf sin-Gordon
model}. We will calculate
$$
\R_2(q)=e^{-i\h{K}(je)}\int
DM(u,j)e^{-iU(u,e)}e^{iS_0(u)}|\Ga(q;u)|^2, $$

The mapping $J:~\{u,p\}(x,t)\rar\{\x,\eta\}(t)$ gives, up to constant
(infinite) coefficient:
$$
DM(u_c,p_c)_N=DM(\x,\eta)
$$$$
=\prod_td\x d\eta\d(\dot{\x} -\f{\pa h_j}{\pa
\eta})\d(\dot{\eta} +\f{\pa h_j}{\pa\x})
$$
$$
h(\eta)=\int dp \s (r)\sqrt{r^2+m^2} +\sum^{N}_{i=1}h(\eta_i),
$$
\be
\hat{K}(e_{\x},e_{\eta};j_{\x},j_{\eta})
=\f{1}{2}Re\int_{C_+} dt \{\hat{j}_{\x} (t)\cdot \hat{e}_{\x}(t)+
\hat{j}_{\eta}(t)\cdot \hat{e}_{\eta}(t)\}.
\ee
$$
U(u_N;e_{\x},e_{\eta})=-\f{2m^2}{\la^2}\int dx dt \sin\la u_N~
(\sin \la e -\la e)
$$
with
$$
e(x,t)=e_{\x}(t) \cdot \f{\pa u_N (x;\x,\eta)}{\pa \eta (t)}-
e_{\eta}(t) \cdot \f{\pa u_N (x;\x, \eta)}{\pa \x (t)}.
$$

One-soliton configuration ($\b =\f{\la^2}{8}$):
$$
u_s=-\f{4}{\la}\arctan\{\exp(mx\cosh\b\eta-\x)\}
$$
Bounded mode:
$$
u_b=-\f{4}{\la}\arctan\{\tan\f{\b\eta_2}{2}
\f{mx\sinh \f{\b\eta_1}{2}\cos \f{\b\eta_2}{2}-\x_2}
{mx\cosh \f{\b\eta_1}{2}\sin \f{\b\eta_2}{2}-\x_1}\}.
$$
Corresponding energies:
$$
h_s(\eta)=\f{m}{\b}\cosh \b\eta,~~~
h_b(\eta)=\f{2m}{\b}\cosh \f{\b\eta_1}{2}\sin\f{\b\eta_2}{2}\geq 0.
$$

Following to {\bf Suggestion B}
$$
 \R_2(q)=0.
$$
Indeed, $v_X(u_c)=sin\{\la u_c\}\pa u_c/\pa X$ and
$$
\{\f{\pa u_c}{\pa X}\}\bigcap\pa_X T^*G=\emptyset
$$

$\bullet$ The mapping of {\bf scalar $O(4,2)$ field theory}
on the $W_O(8)=T^*\bar{G}_1(4)\times R(5)$ space gives $\R_2(q)\neq0$
since $\{\vp\}\bigcap{\rm inf}\pa_{\eta_1}W_O\neq\emptyset,$ where
$$
\vp(x)=(\f{4}{g\eta_1^2})^{1/2}\{ (1+\f{(x-x_0)^2}
{\eta_1^2})^2
+ (2\f{\eta_2l_\mu (x-x_0)^\mu}{\eta_1})^2\}^{-1/2}
=O(1/\sqrt g),
$$
where:
$
l_\mu l^\mu=+1/\eta_2^2\geq0,~~\vec{l}^2=1.
$
{\it Note,} other directions in the $W_O$ space did not give
contributions

One should use:
$$
U(\vp,e)=2g\Re\int_{C_+}d^3x dt\vp(x)e^3(x)
$$
$$
DM(\x, \eta)=d^3x_0 d^3l
\d(\vec{l}^2-1)dt_0 \d(\x_1(0)-\x_2(0))
$$$$
\times\prod_t d^2\x (t)d^2\eta (t) \d^2(\dot{\x}-\f{\pa {h}_j}{\pa \eta})
\d^2(\dot{\eta}+\f{\pa {h}_j}{\pa \x})
$$$$
\h{K}_t (j,e)=\f{1}{2}\int dt (\h{j}_\x\cdot\h{e}_\x
+\h{j}_\eta\cdot\h{e}_\eta)
$$

$Note$, by definition, $\R_2(q)\sim\d(q^2-m^2),$ where $m=h_r\neq0$ is
the renormalized energy of $u_c$, $h_r=h(u_c)+O(\hbar)$, $h(u_c)\sim
1/\eta_1$ (!) by definition.

$\bullet$ The mapping of {\bf Yang-Mills theory} on $W_O\t G$ gives:
$$
DM(\x,\eta,\La)=d^3x(0)d^3l\d(l^2-1)\d(\x_{10}-\x_{20})
$$$$
\prod_{x,t;a}d^2\x(t)d^2\eta(t)d\La_a(x,t)
%$$$$
\d(\dot{\x}-\o-j_\x) \d(\dot{\eta}-j_\eta)
$$
$$
\h{K}(je)=\f{1}{2}\int dt\{\h{\bf j}^a_\x\cdot\h{\bf e}^a_\x+
\h{\bf j}^a_\eta\cdot\h{\bf e}^a_\eta\}
$$
$$
U({\bf u},\bar{\bf e})=S_o({\bf u})-\int d^4x\prod_a\le\{\bar{\bf
e}_a\cdot\f{\pa }{\pa {\bf u}_a}\ri\}v({\bf u}),
$$
$$
\bar{\bf e}_a={\bf e}_\x\f{\pa {\bf u}_a}{\pa\eta}-
{\bf e}_\eta\f{\pa {\bf u}_a}{\pa\x}.
$$
$Note$, measure is ghost fields free.

\end{document}